%% file: KTAP-arXiv-18-12-2020.tex
\documentclass[12pt,reqno]{amsart}
\usepackage[utf8]{inputenc}
\usepackage{epsfig}
\usepackage{colortbl}
\usepackage{ulem}
\everymath{\displaystyle}
\usepackage{geometry}
\usepackage{fancyhdr}
\usepackage{tikz}
\usetikzlibrary{shapes,arrows}

\usepackage{multirow}

\usepackage{hyperref}

\usetikzlibrary{shapes,arrows,decorations.markings}
\usetikzlibrary{decorations.pathmorphing}

\tikzstyle{block_long} = [rectangle, draw, fill=blue!20,
    text width=10.0em, text centered, rounded corners, minimum height=3em]
\tikzstyle{block_medium} = [rectangle, draw, fill=blue!20,
    text width=6.0em, text centered, rounded corners, minimum height=3em]
\tikzstyle{block_short} = [rectangle, draw, fill=blue!20,
    text width=3.0em, text centered, rounded corners, minimum height=3em]
\tikzstyle{line} = [thick, draw, dashed,  -stealth']

\def\@xthm#1#2{\@beginassumption{#2}{\csname the#1\endcsname}{}\ignorespaces}
\def\@ythm#1#2[#3]{\@opargbeginassumption{#2}{\csname the#1\endcsname}{#3}\ignorespaces}%
\def\@beginassumption#1#2#3{\par\addvspace{8pt plus3pt minus2pt}%
              \noindent{\csname#1headfont\endcsname#1\ \ignorespaces#3 #2.}%
              \csname#1font\endcsname\hskip.5em\ignorespaces}
\def\@endassumption{\par\addvspace{8pt plus3pt minus2pt}\@endparenv}
\usepackage{mathrsfs,bbm}
\usepackage[american]{babel}

\newcommand{\f}{f}

\newcommand{\no}{\noindent}

\newcommand{\vf}{\varphi}

\newcommand{\bomega}{\boldsymbol{\omega}}

\newcommand{\bx}{{\boldsymbol{x}}}

\newcommand{\bv}{{\boldsymbol{v}}}
\newcommand{\bu}{{\boldsymbol{u}}}
\newcommand{\bw}{{\boldsymbol{w}}}
\newcommand{\bz}{{\boldsymbol{z}}}

\def\va{\raise 2pt\hbox{,}}

\def\cL{\mathcal{L}}

\def\cD{\mathcal{D}}
\def\cC{\mathcal{C}}

\def\cP{\mathcal{P}}

\newcommand{\p}{\partial}

\font\dc=cmbxti10 

\input epsf
\begin{document}


\markboth{N.~Bellomo, D.~Burini, G.~Dosi, L.~Gibelli, D.~Knopoff, N.~Outada, P.~Terna, M.E.~Virgillito}{What is life for mathematical sciences}

%
%

\title{Active particles methods  \\towards modeling in science and society}

 \vspace{1cm} 

\author{Nicola Bellomo$^{(1)}$, Diletta Burini$^{(2)}$, Giovanni Dosi$^{(3)}$,  Livio Gibelli$^{(4)}$,  Damian Knopoff$^{(5)}$, Nisrine Outada$^{(6)}$, Pietro Terna$^{(7)}$, and Maria Enrica Virgillito$^{(8)}$} 

\address{$^{(1)}$\tiny{University of Granada, Spain.}  {\tt nicola.bellomo@polito.it}}

\address{$^{(2)}$\tiny{University of Perugia, Italy.} \\ {\tt dilettaburini@alice.it}}

\address{$^{(3)}$\tiny{Scuola Superiore Sant'Anna, Pisa Italy.} \\ {\tt giovanni.dosi@santannapisa.it}}

\address{$^{(4)}$\tiny{School of Engineering, Inst. for Multiscale Thermofluids\\ University of Edinburgh, U.K} \\ {\tt livio.gibelli@ed.ac.uk}}

\address{$^{(5)}$\tiny{BCAM, Basque Center for Applied Mathematics, Spain, and Centro de Investigaci\'on y Estudios de Matem\'atica (CONICET) and Famaf (UNC), C\'ordoba, Argentina.}\\ {\tt damian.knopoff@unc.edu.ar}}

\address{$^{(6)}$\tiny{Cadi Ayyad University, Faculty of Sciences Semlalia, LMDP, Morocco and UMMISCO (IRD-Sorbonne University, France)} {\tt outada@ljll.math.upmc.fr}}

\address{$^{(7)}$\tiny{University of Torino, Torino, Italy, and Fondazione Collegio Carlo Alberto, Torino, Italy.}\\ {\tt pietro.terna@unito.it}}

\address{$^{(8)}$\tiny{Scuola Superiore Sant'Anna, Pisa, Italy.} \\ {\tt mariaenrica.virgillito@santannapisa.it}}

\maketitle



\begin{abstract}
 This paper is a first step to chase  the ambitious objective of developing a mathematical theory of living systems. The contents refer modeling large systems of interacting living entities with the aim of describing their collective behaviors by differential models. The contents  is in three parts. Firstly, we  derive the mathematical method;  subsequently, we show how the method can be applied in a number of case studies related to well defined living systems and finally, we look ahead to research perspectives focusing both on mathematical methods and further applications. 
\end{abstract}

\begin{center}
\textit{To the memory of Pietro Greco - 20-04-1955 --- 18-12-2020}
\end{center}

\section{Aims and plan of the paper}\label{sec:1}

The search for a mathematical theory of living systems is one of the great scientific challenges of this century. Indeed, scientists are waiting for
a rigorous formalization, by mathematical equations and computer architectures, of living systems in biology, economy, social sciences, and various disciplines within the general framework of the so-called \textit{soft sciences}. This challenging objective might even appear  too ambitious as it goes far beyond the present state of the art and it requires new mathematical tools, actually a new mathematical theory.

The specific objective we are chasing requires not only contrasting the negativist position which claims that some research fields should be based only on heuristic methods without deserving the use of the adjective \textit{scientific}, but also  going beyond the classification into hard and soft sciences focusing on the ranking proposed by August Comte~\cite{[COMTE]}. The main concept proposed in~\cite{[COMTE]} is that a classification can be proposed, where physics and chemistry are the hardest sciences, sociology belongs to the softest, while biology is in the middle. An immediate question follows: \textit{How mathematics can be classified?} The reply  is given in Chapter 7 of~\cite{[BBGO17]}, where the authors suggest to replace the definition  \textit{soft sciences}  with the following: {\dc Science of Living Systems}.
This vision gives to mathematics an essential role towards a unified vision of all sciences which goes beyond any classification from soft to hard.
Indeed, a first step to develop a strategy to take into account that in the case of the living matter the approach cannot be supported by field theory~\cite{[BBGO17]}. The strategy consists in replacing the field theory  by a mathematical structure (say a mathematical theory) suitable to capture, as far as it is possible, the complexity features of living systems. This structure defines the conceptual framework for the derivation of models in different fields of soft sciences. Applications refer  to a broad variety of possible fields, for instance  modeling of social dynamics, financial markets, dynamics of multicellular systems, immune competition, individual and collective learning, and the modeling of large systems of self-propelled particles such as crowds and swarms. The collection of essays edited in~\cite{[Ball12]} reports about a broad variety of applications within the concept that we do live in a complex environment.

This paper is devoted to the presentation and critical analysis of some recent results devoted to  mathematical tools which claim to describe some important features of living systems constituted by several interacting entities. The presentation is not limited to theory, as applications play an important role on the overall contents due to their contribution to enlighten the bridge between mathematics and real systems. In addition, a critical analysis is a key feature of each section looking ahead  to further development of the theory.

 The first part of our paper is devoted to the derivation of mathematical tools in view of a mathematical theory which includes some new ideas, with respect to those proposed in~\cite{[BBGO17]} and a more general vision of active particle methods. Subsequently, we present a selection of applications  focused on the modeling of dynamics of living systems which can be modeled by the theoretical tools presented in the first part of the paper. The selection accounts for different features of models with special attention to the role  of space dynamics by showing how space can have an influence over the collective behavior of the whole system.

 Finally, we look ahead to research perspectives within the framework of the complex dialogue between soft and hard sciences. Each section is closed by a critical (and self-critical) analysis whose main aim consists in enlightening how far the mathematical tools proposed in our paper succeed in chasing the mythical  objective of designing a mathematical theory of living systems.

 This paper has been written during the \textit{COVID-19} outbreak and the subsequent pandemic, which has brought to almost all countries across the globe huge problems affecting health, safety, economics, and practically all expressions of collective behaviors in our societies. One of the modifications in our ways of thinking and communicating is the loss of the vis-a-vis way of developing dialogues. Lack of direct dialogue has definitely created additional difficulties to young researchers. Accounting for this difficulty, we have produced training course of seven Lectures~\cite{[BBDG20]} which is offered to young researchers, motivated by lack of communication, as well as by the present difficulty to move across research centers which is, indeed, an important engine of scientific growth.

\vskip.1cm \noindent Section 2 is devoted to design a strategy towards the derivation of mathematical tools to model large systems of living interacting entities, where their collective dynamics is generated by interactions among the said entities and the external environment. The strategy essentially consists in selecting the key complexity features of living systems, in general, to be specialized into the class of systems object of the modeling approach. 

\vskip.1cm \noindent In Section 3 we transfer the strategy proposed in Section 2 into mathematical structures which provide the conceptual framework for the derivation of models which are obtained by inserting in the structures the mathematical description of interactions. The said structures, which is derived within the general framework of the so-called \textit{kinetic theory of active particles}, is deemed to take the place of the field theories available to support the derivation of models for physical systems of the inert matter. This section presents also some reasonings on the links between the micro-scale (individual based)  and the mesoscopic scale (kinetic).

\vskip.1cm \noindent In Section 4 we tackle  one of the various aspects of a multiscale vision of living systems and shows how the mathematical theory presented in Section 3 can be developed, according to analogous physical  assumptions, within the framework of dynamical systems corresponding to
an individual based description.  This approach is leads to the so-called \textit{mathematical theory of behavioral swarms} introduced in~\cite{[BHO20]} and applied to modeling the dynamics of prices in~\cite{[KTS20]}.

\vskip.1cm \noindent Section 5 presents a concise survey of applications of the mathematical theory presented in the preceding section. The survey refers to both approaches, namely the kinetic theory of active particles and the theory of swarms made of active particles. The survey is limited to very recent years with the aim of enlightening new ideas on the modeling of interactions.

\vskip.1cm \noindent Finally, we present in the last section a  critical analysis mainly focused on multiscale methods to be developed as the key strategy  to  further modeling perspectives.

\section{A strategy towards modeling living systems}\label{sec:2}

This section is devoted to the design of a general strategy towards the modeling of the collective dynamics of large systems of interacting living entities. The quest towards this challenging objective is in three steps, each of them treated in the following subsections. Firstly, we present a general conceptual-philosophical framework,  subsequently a strategy is proposed in view of mathematical formalizations, lastly a critical analysis is proposed, focusing on a suitable interpretation of the scaling problem, to make operative the strategy towards an appropriate selection of mathematical tools.

\subsection{Conceptual framework}\label{subsec:2.1}

A conceptual framework is proposed here to provide a support to the derivation of the modeling strategy based also on some scientific  works  selected according to the authors' bias accounting  for their contribution to understand the complex interactions between mathematical sciences and the dynamics of living systems.

\vskip.2cm   \noindent  $\bullet$   \textbf{Immanuel Kant  1724--1804}, \textit{Living Systems: Special structures organized and with the ability to chase a purpose~\cite{[KANT]}.}

\vskip.2cm   \noindent  $\bullet$   \textbf{Erwin Schr\"odinger  (1887--1961)}  looked for a physical theory, where cells modify their state due to interactions with other cells~\cite{[ES1944]}. Schr\"odinger's pioneering ideas chased a systems approach motivated by the study of mutations (some of them also induced by external actions such as radiations). We can argue that one of his intuitions was that the dynamics at the level of cells is driven by the dynamics at the molecular scale. This concept is nowadays the most important hint of the interactions between mathematics and biology, where understanding the link between the dynamics at the molecular scale of genes and the functions expressed at the level of cells is the key towards the possible derivation of a bio-mathematical theory. The following sentence:
\begin{quote}
\textit{Living systems have the ability to extract entropy to keep their own at low levels}
\end{quote}
identifies ability  of living systems to develop their own strategy. Hence the concept of \textit{active particles} was already introduced.

\vskip.2truecm \noindent   $\bullet$   \textbf{Lee Hartwell (born 1938)}, Nobel Laureate in 2001, firmly indicates~\cite{[HART99]} that the mathematical approach to the description of the dynamics of the inert matter cannot be straightforwardly applied to living systems:
\begin{quote}
\textit{Biological systems are very different from the physical or chemical systems of the inanimate matter. In fact, although living systems obey the laws of physics and chemistry, the notion of function or purpose differentiate biology from other natural sciences. Indeed, cells are not molecules, but have a living dynamics induced by the lower scale of genes and is organized into organs.}
\end{quote}

This statement directly looks forward to a challenging research perspective whose first step consists in  acknowledging that the mathematics used for the inert matter fails  when applied to the living matter.
\begin{figure}
\begin{center}
\includegraphics[width=0.25\textwidth]{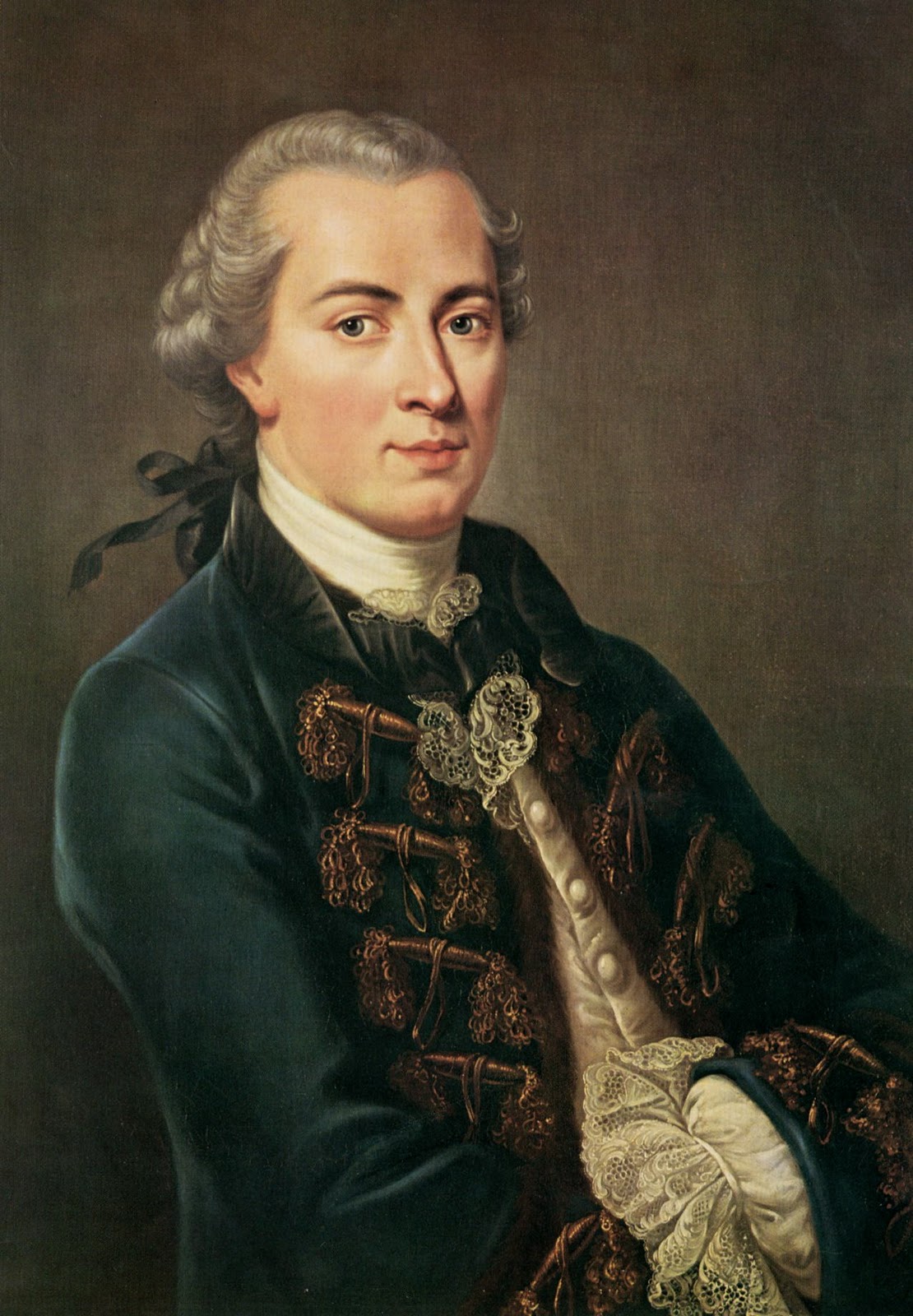} \hspace*{1cm} \includegraphics[width=0.24\textwidth]{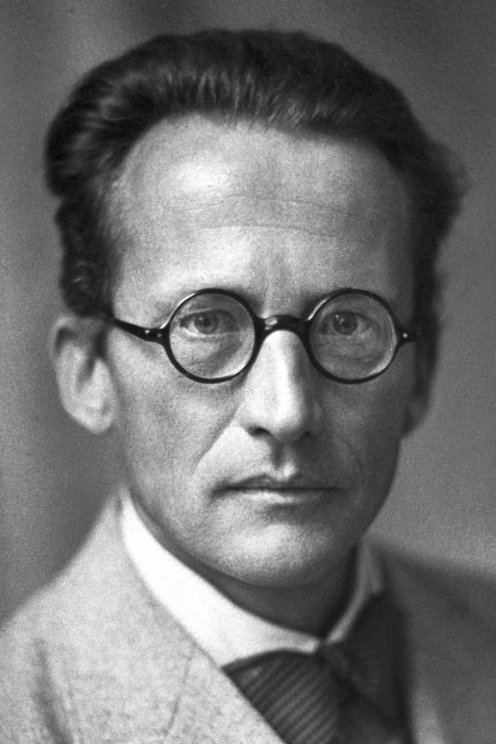} \hspace*{1cm} \includegraphics[width=0.24\textwidth]{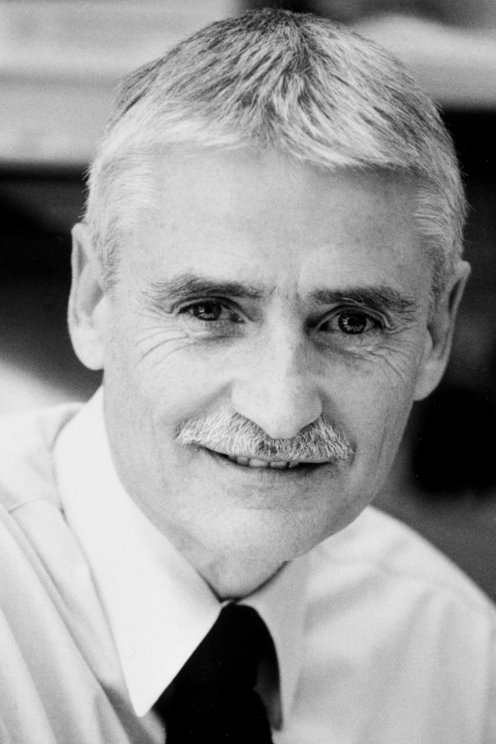}
\end{center}
\caption{Kant - Schrodinger - Hartwell}
\end{figure}

\vskip.2truecm Following these two milestones, some specific models have been proposed to describe the dynamics of the aforementioned class of dynamical systems by methods somehow inspired by the mathematical kinetic theory and by the Boltzmann equation in particular~\cite{[CIP93]}. For instance, Prigogine and Hermann developed an approach to  describe the dynamics of vehicular traffic on highways~\cite{[PH71]}. An important feature of this model is that the car-driver subsystem is viewed as an active particle whose ability is heterogeneously distributed, while the overall state of the system is defined by a probability distribution over the micro-scale state. A deep study of the heterogeneity properties has been clearly identified by the structures proposed in~\cite{[PF75]}.

Methods from the kinetic theory, have been subsequently developed by various authors, for instance on the modeling of the social dynamics of families of insects in~\cite{[JS92]}, or of the immune competition between cancer and immune cells~\cite{[BF94]}. These pioneer papers have been followed by a vast literature mainly developed in this century. A review and a critical analysis on kinetic theory methods is postponed to the next section specifically devoted to transfer the concepts of this section into a mathematical theory. Here we simply indicate the sharp critical analysis
presented in~\cite{[ARI19]} which enlightens the conceptual differences between the classical kinetic theory and that referred to active particles.

In addition, one has to account that all living systems are evolutionary by a dynamics followed by selection. This pseudo-darwinian  feature does not belong, as we shall see, to biology only, but it characterizes all living, hence behavioral, hence complex, systems, for instance in economy and sociology, where the dynamics, driven by learning, includes mutations and selection.

The awareness of Hartwell's legacy motivates a quest towards the search of a rationale  to chase the objective of the derivation of a mathematical theory of living systems by going far beyond the classical methods valid for the inert matter. A preliminary contribution to this challenging objective is delivered by the answer to three key questions presented in the following.

\vskip.2truecm \noindent  $\bullet$   \textbf{KQ1: What is complexity?} \textit{Complexity viewed as a barrier}. Imagine a world correctly described in mathematical and logical terms. That would make realistic the Leibniz dream (free translation from Latin language):

\begin{quote}
\textit{In the future, when an issue is controversial, it will not be necessary to dispute between two philosophers but between two subjects able in computations. It will suffice them to keep the abacus into their hands, sit down, and say each other---in a friendly way)---start making calculations},~\cite{[LEIBNIZ]}.
\end{quote}

Unfortunately (or luckily?), two hard constraints preclude the actual achievement of a world so entirely known to allow such exciting solutions.

\begin{itemize}

\item	The so-called chaos, due to the presence of dynamical systems, whose apparently-random states of disorder and irregularities are often governed by deterministic laws (that are highly sensitive to initial conditions);

\vskip.1cm \item	The phenomenon of complexity, due to the emergence of entirely new properties at any new level of aggregation.

\end{itemize}

Arguably, Leibnitz was not aware of the existence of  them. Therefore, we might pose to ourselves the following question: \textit{from where to start to understand complexity?} In Aristotle, complexity seems opposed to simplicity as a matter of lifestyle. In Latin, the world \textit{complexus} means what is woven together.

In the 40s of the last century, von Neumann was working with automata and their complexity, but:  \textit{he described his own concept of complexity as ''vague, unscientific and imperfect''} (from McMullin, 2000).

 If we jump to the 60s of the last century, we have the Kolmogorov complexity, as the length of the shortest computational sequence producing the object as output. Beautiful, but again it is not a reply to our search about what complexity is.
 The concept was there, but missing a clear interpretation and definition, confused with the a-scientific and anti-reductionist holism, i.e., the idea that we should view many systems (physical, biological, social, our body, etc.) as wholes, not merely as collections of parts. Sure, but then what?
So, neither holism nor simple reductionism,  but with Nobel Laureate Philip Anderson (born 1923), in 1972 the ``More is different'' clarification~\cite{[AND72]}:

\begin{quote}
 \textit{(p.393) The reductionist hypothesis may still be a topic for controversy among philosophers, but among the great majority of active scientists I think it is accepted without questions. The workings of our minds and bodies, and of all the animate or inanimate matter of which we have any detailed knowledge, are assumed to be controlled by the same set of fundamental laws (\ldots)  The main fallacy in this kind of thinking is that the reductionist hypothesis does not by any means imply a ``constructionist'' one (\ldots) The constructionist hypothesis breaks down when confronted with the twin difficulties of scale and complexity. The behavior of large and complex aggregates of elementary particles, it turns out, is not to be understood in terms of a simple extrapolation of the properties of a few particles. Instead, at each level of complexity entirely new properties appear (\ldots).}
\end{quote}

Often, complexity is related to biological systems, however this is not an exhaustive vision as complexity is everywhere in our world~\cite{[Ball12]}. In particular, the focus of our paper goes to evolutionary economy referring to the conceptual framework in~\cite{[DV17]}, where it is given evidence of the role of complexity in economical systems. The following quotation from Nobel Laureate in economy Herbert Simon  has been extracted  from~\cite{[SIMON69]} and reported in~\cite{[DV17]} to enlighten the initial step towards linking economy to the theory of complexity.

 \begin{quote}
 \textit{Roughly by a complex system I mean one made up of a large number of parts that interact in a non-simple way. In such systems, the whole is more than the sum of the parts, not in an ultimate metaphysical sense, but in the important pragmatic sense that, given the properties of the parts and the laws of their interaction, it is not a trivial matter to infer the properties of the whole.}
\end{quote}

 At this end, as a key milestone towards the development of mathematical theory, it is necessary   transferring the aforementioned general concepts to an assessment of the relevant complexity features by answering to the second key question.

\vskip.2truecm \noindent \textbf{KQ2: Which are the main complexity features of living system?} A mathematical theory of living system should arguably attempt to capture  the complexity feature of living systems~\cite{[BBGO17]}. Therefore, the answer to this question aims at contributing to the key objective of our paper. Without naively claiming that our reply can be exhaustive, our proposal for a selection of five key features is as follows:

\vskip.1cm \noindent {\dc 1. Ability to express a strategy:} Living entities are capable to develop
specific {\sl strategies} and {\sl organization abilities} that depend on the state  of the surrounding environment.

\vskip.1cm \noindent  {\dc  2. Heterogeneity:} The ability to express a strategy is not
the same for all entities as {\sl expression of heterogeneous behaviors} is a common feature of a  great part of living systems.

\vskip.1cm \noindent   {\dc 3. Nonlinearity of  interactions:} Interactions are nonlinearly additive as well as nonlocal as immediate neighbors, but in some cases also distant entities, are involved.

  \vskip.1cm \noindent {\dc 4. Learning ability:} Living systems receive inputs  from their environments and have the ability to learn from past experience.

 \vskip.1cm \noindent {\dc 5. Darwinian mutations and selection:} All living systems are evolutionary, as birth processes can generate  entities more fitted to the environment, who in turn generate new  entities again more fitted to the outer environment.

\vskip.1truecm \noindent \textbf{KQ3: What is the black swan?} The expression \textit{black swan} has been introduced to denote  unpredictable events which are far away from those generally observed by repeated empirical evidence. Let us report the definition by Taleb~\cite{[TALEB]}:
\begin{quote}
\textit{A Black Swan is a highly improbable event with three principal  characteristics: It is unpredictable; it carries a massive impact; and, after the fact, we concoct an explanation that makes it appear less random, and more predictable, than it was.}
\end{quote}

Actually, the concept of the \textit{black swan} is associated to the concept of \textit{unpredictable event} or, negatively \textit{not predicted event}, but we wish stressing that we want to refer this concept to the ability of mathematical models which should provide all possible scenarios including events which are not those one can figure out. This vision has a well defined implication on the modeling approach which should not include any relaxation term inserted in the model based on observations by empirical data. Indeed, this is a key issue towards the derivation of a mathematical theory of living systems to be carefully tackled in the following.

\subsection{From philosophical thoughts to figurative fantasy}\label{subsec:2.2}

Let us now leave the various concepts presented until now and give some space to our fantasy. A free interpretation in\cite{[BKS13]} suggests that the
\textit{Metamorphosis-III} by Cornelius Escher:
 \begin{center}
$https://arthive.com/it/escher/works/200075~Metamorphosis$
\end{center}
depicts most of the aforementioned features, for instance a strategy to transform a village with houses with almost uniform shapes into an heterogeneous village which includes architectures with different shapes.

We can observe that the evolution is selective as shown by the transition from essential shapes to an organized village, where all available spaces are well exploited. The presence of a church, that takes an important part of the space and a somehow central position, indicates the presence of a cultural evolution. This might even reflect a multiscale dynamics. In fact, it results as the output of the action from the micro-scale of individuals to the macro-scale of the village. In addition, the last part of Metamorphosis-III shows a sudden change from a peaceful village to a chess plate which represents a battle between two antagonist armies. If we hide this part, we should admit that it is a sudden change which is not predicted by early signal. The third key question specifically refers to this topic.

 Let us return to Cornelis Escher, by noticing that the village exists in reality (it is in the Mediterranean  coast immediately on the South of the village of Amalfi), but the following question can be posed: \textit{Does the Tower truly exists?}

\begin{figure}[t!]
\begin{center}  \label{Atrani}
  \includegraphics[height=5cm, width=8.0cm]{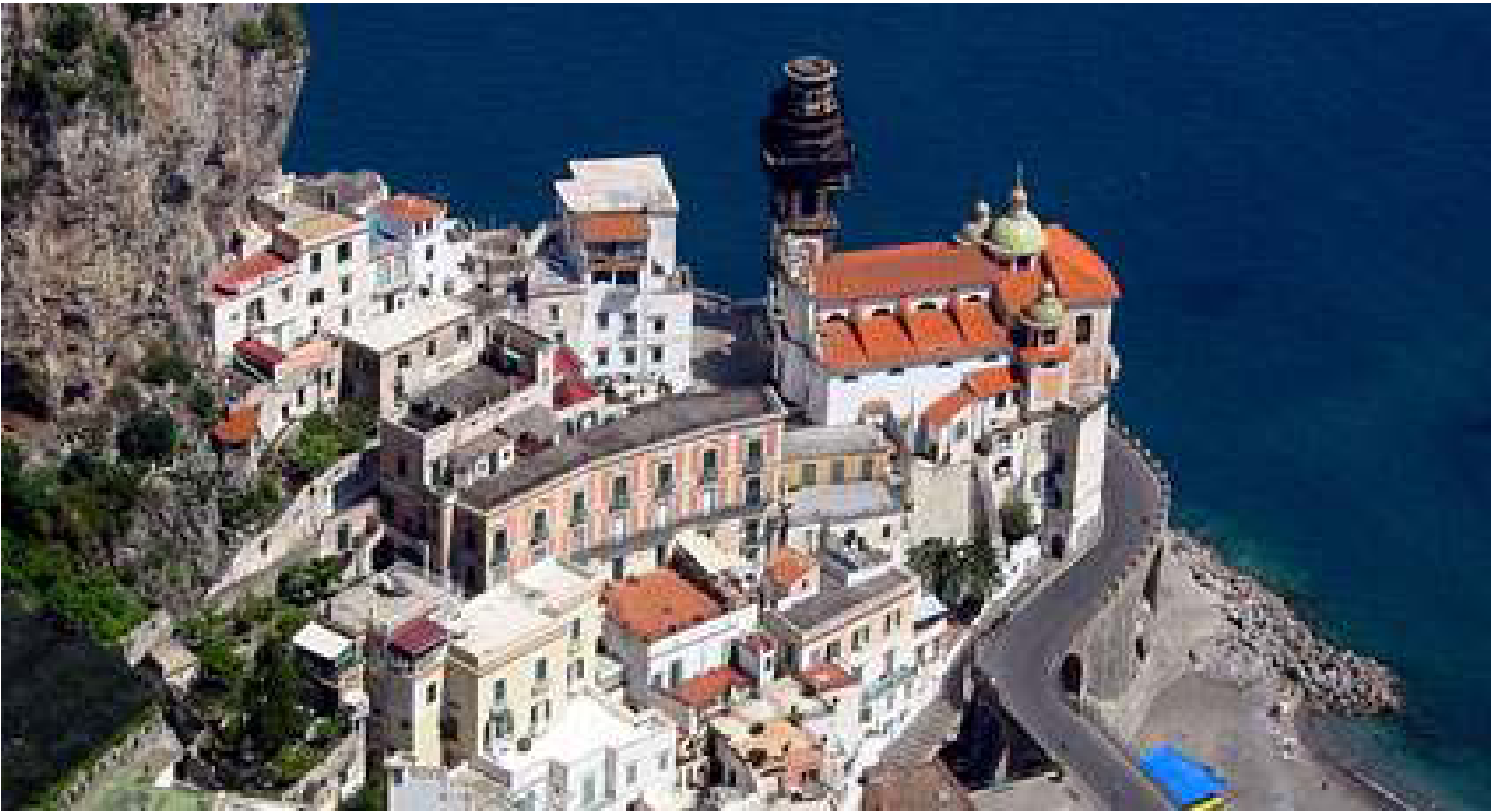}
    \includegraphics[height=5cm, width=7.0cm]{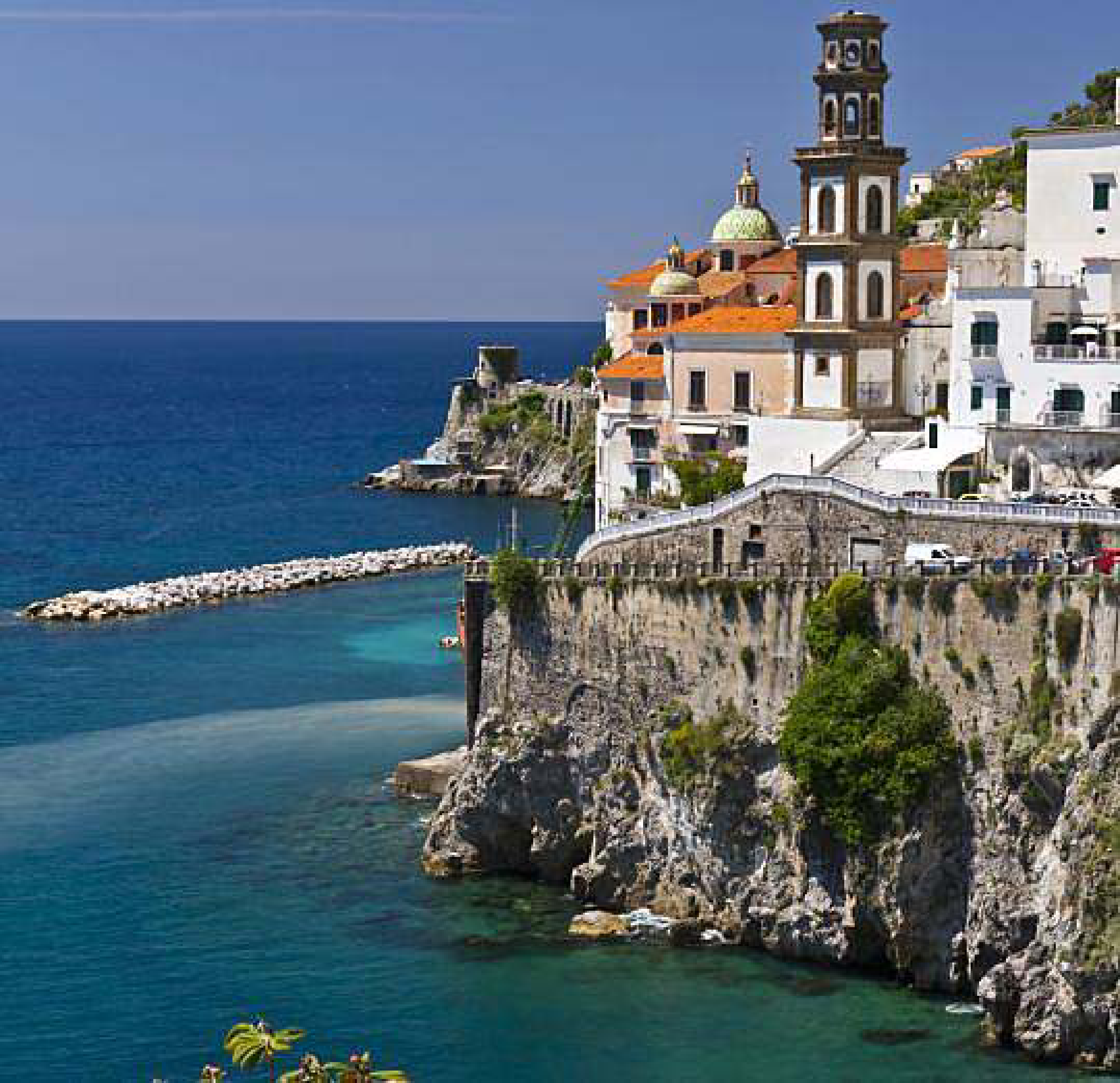}
\caption{Different visions of of the village Atrani}
\end{center}
\end{figure}

 The answer is that the real village looks at the sea, while the tower cannot be observed looking at it from right to left by an observer looking at the sea from the village, as shown on the left picture of Figure~\ref{Atrani}.  But an observer   the rear left of the village can observe a tower on the cape on the right of the village. In reality, the observation goes from south to north, as shown on the right picture of Figure~\ref{Atrani}.

 So far the Metamorphosis teaches us that different visions of the same object can be represented into one unified collective representation. Indeed this is one of the specific features of complex systems. If we leave some additional freedom to our fantasy, the tower can be interpreted as an early signal that an extreme event is going to happen. The various changes in the picture can be interpreted as predictable emerging behaviors, while the last one appears as a non-predictable event.  Escher has gone through the experience of two world wars, where  peaceful villages  transformed into a battlefield between the two armies of the chess plate.

\subsection{On a strategy towards modeling living systems}\label{subsec:2.3}

 We propose a modeling strategy  accounting that, in the case of the living matter, the approach cannot be supported by field theory which explains why the definition \textit{soft sciences}  with \textit{science of living systems}.  The strategy consists in replacing the field theory  by a mathematical structure (say mathematical theory) suitable to capture, as far as it is possible, the complexity features of living systems. This structure defines the conceptual framework for the derivation of models which are obtained by inserting models of interactions into the structure itself. We refer our approach to large systems of interacting living entities. In more details, the sequential steps of the strategy can be summarized as follows:
 \begin{enumerate}

\item  \textit{Understanding the links} between the dynamics of living systems and their \textit{complexity features};

\vskip.1cm  \item  \textit{Derivation of a general mathematical structure}, consistent with the aforesaid features, with the aim of offering the
conceptual framework toward the derivation of specific models;

\vskip.1cm  \item  \textit{Design of specific  models} corresponding to well defined classes of systems by implementing the said structure with
suitable models of  individual-based, micro-scale, interactions;

\vskip.1cm  \item  \textit{Validation of models} by quantitative comparison of the dynamics
predicted by them with that one delivered by empirical data. Models are required to reproduce qualitatively emerging behaviors.

\end{enumerate}

\begin{figure}[t!]
\begin{center}
\begin{tikzpicture}[node distance = 2cm, auto]
 \node [block_long] (crowd) {Phenomenological interpretation};
\node [block_long, below of=crowd] (complex) {Complexity features\\of living systems};
 \node [block_long, below of=complex] (model) {Mathematical structures};
 \node [block_medium, left of=model,xshift=-2cm] (learning) {Interaction dynamics};
 \node [block_medium, right of=model,xshift=2cm] (social) {External actions};
 \node [block_medium, below of=model,xshift=-4cm] (mathematical) { Mathematical topics };
 \node [block_long, below of=model] (validation) {Derivation of models};
 \node [block_medium, below of=model,xshift=4cm] (simulation)  {Computing \\simulations};
\node [block_long, below of=validation] (artificial){Model validation};
 \draw [->] (crowd.south) -- (complex.north);
 \draw [->] (complex.south) -- (model.north);
 \draw [->] (learning.east) -- (model.west);
 \draw [->] ([yshift=0.3cm] validation.west) -- ([yshift=0.3cm] mathematical.east);
 \draw [->] ([yshift=0.3cm] validation.east) -- ([yshift=0.3cm] simulation.west);
 \draw [->] ([yshift=-0.3cm] mathematical.east) -- ([yshift=-0.3cm] validation.west);
 \draw [->] ([yshift=-0.3cm] simulation.west) -- ([yshift=-0.3cm] validation.east);
 \draw [->] (social.west) -- (model.east);
 \draw[->] ([xshift=-1cm] model.south) -- (mathematical.north);
 \draw[->] (model.south) -- (validation.north);
 \draw[->] ([xshift=1cm] model.south) -- (simulation.north);
 \draw[->] ([xshift=-0.5cm] artificial.north) -- ([xshift=-0.5cm] validation.south);
\draw[->] ([xshift=0.5cm] validation.south) -- ([xshift=0.5cm] artificial.north);
\end{tikzpicture}
\end{center}
\caption{Modeling strategy}
\label{rationale}
\end{figure}
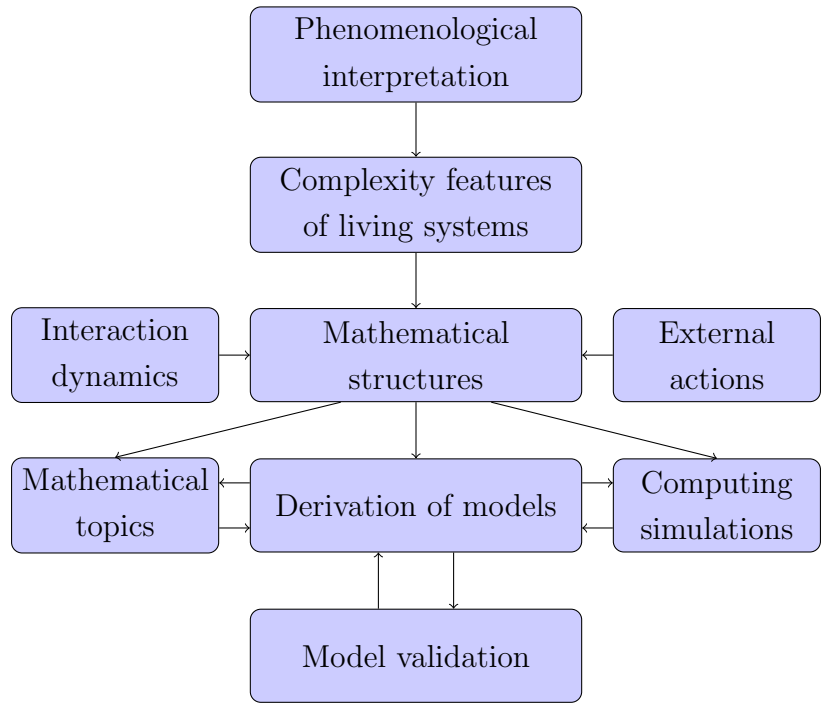

This strategy, which leads to a modeling rationale, is represented in figure~\ref{rationale} which indicates, by a flow-chart, how the observation of the real system moves to models, which only approximate, physical reality and, consequently, need to be validated.

\noindent Some additional remarks contribute to further enlightening  the flow chart.

\vskip.1truecm \noindent  $\bullet$  \textit{Multiscale aspects:} Modeling always needs a {\sl multiscale approach}, where the dynamics at the large scale needs to be properly related to the dynamics at the low scales. For instance, the functions expressed by  a cell are determined by the dynamics at the molecular (genetic) level.

\vskip.1truecm   \noindent $\bullet$  \textit{Role of the environment:} The environment  evolves in time, in several cases also due to interactions with the internal living system.

\vskip.1truecm  \noindent  $\bullet$  \textit{Large deviations:} Emerging behaviors often present large deviations although
the qualitative behaviors is reproduced. In this case, small deviations in the input create large deviations in the
output.

\vskip.1truecm  \noindent  $\bullet$  \textit{Individuals within a certain FS can aggregate into groups of affinity:} Communications  and subsequent dynamics can take advantage (or disadvantage) from the said aggregation  by creating a new communication network.

\subsection{The scaling problem}\label{subsec:2.4}

 As it is known, the representation and modeling of dynamical systems can be developed at each of the three scales namely \textit{microscopic} (individual based), \textit{macroscopic} (hydrodynamical), and at the intermediate \textit{mesoscopic} (kinetic) scale. In the kinetic theory approach, the dependent variable is a distribution function over the microscopic state of the individuals.

Our study, refers to the collective dynamics of several heterogeneous interacting individuals. Heterogeneity, which affects interactions, motivates the selection of the kinetic theory approach as the most appropriate towards modeling. In fact, it can naturally account for heterogeneity and stochastic interactions~\cite{[BBGO17]}. On the other hand, the number of interacting entities is not, in most cases, large enough to justify continuity assumption of the aforementioned distribution function. This key difficulty cannot be hidden and has to be carefully treated in the derivation of the mathematical structure in addition to further developments of  the methods of classical models of the  kinetic theory  which cannot be straightforwardly applied to living systems.

For instance, the celebrated Boltzmann equation is based on the assumption of a rarefied flow where only binary, short range, interactions occur, while interactions in the Vlasov equations are distributed in the whole space which is not the case of living systems, where the domain of interactions refers both to sensitivity and visibility of individuals. In addition, interactions of classical particles preserve mechanical variables, namely mass, momentum, and energy, while these properties are often lost in the case of living systems.

The mathematical theory reported in the next section is grounded on methods of the kinetic theory consistently with the strategy proposed in this section, however, accounting for the necessary developments imposed by the complexity features of living systems. This selection should be critically examined as none of the scales, standing alone, is sufficient to depict the dynamics of the class of systems under consideration. Indeed, a multiscale vision is necessary and it is a key feature of a mathematical theory of living systems. In addition, the kinetic theory approach requires that the number of interacting entities is large enough to justify the continuity assumption of the distribution function representing the overall state of the system. Therefore, various reasonings on the  multiscale vision are going to be a constant presence in the next sections, while a conceptual alternative is developed in Section 4.

\section{Towards a mathematical kinetic theory of living systems}\label{sec:3}

We show, in this section, how the strategy proposed in Section 2 can be transferred into a \textit{mathematical theory}, where this term is used to refer to mathematical structures suitable to capture, as far as it is possible, the complexity features of living systems. The contents  are focused on the kinetic theory of active particles, where individual entities, called \textit{active particles} (in short a-particles), interact across networked populations. The  micro-scale state of a-particles  includes, in addition to mechanical variables, also a variable, called \textit{activity}, which models the behavioral ability of the individual entities.

As mentioned, there exists a well settled literature on this topic which accounts for those we have indicated as pioneering research works~\cite{[BF94],[JS92],[PF75],[PH71]}. We specifically refer to the book~\cite{[BBGO17]} and, in addition, we  include some recent developments mainly induced by specific applications. In more details on the contents of this section, the theory is presented in Subsection 3.1, while a critical analysis is proposed in Subsection 3.2 to enlighten how far the theory is consistent with the conceptual approach of Section 2 and how further research activity should be developed in the search of alternative methods.

\subsection{The mathematical theory of active particles}\label{subsec:3.1}
We consider the collective dynamics of large systems of interacting individuals. Therefore the kinetic theory approach appears to be the most appropriate to be selected without hiding the key difficulty consisting in that the number of interacting entities is not large enough to fully justify the continuity assumption of the distribution function.

Living entities,  at each interaction, play a game with an output that depends on their strategy often related to surviving and adaptation abilities.
Interactions are modeled by theoretical tools of stochastic game theory which are characterized as follows:

\begin{itemize}

\vskip.1cm \item Stochastic game theory deals with entire populations of players, where strategies with higher payoff might spread over each population by learning related to the individual based and between each individual and the collectivity interactions.

\vskip.1cm \item The strategy expressed by individuals, i.e. active particles, is heterogeneously distributed over the micro-states of players which include both mechanical and activity variables.

\vskip.1cm \item Players are modeled as stochastic variables linked to a distribution function over the micro-state. The pay-off is heterogeneously distributed over players as well,  and it might be motivated by ``rational'' or even ``irrational'' strategies.

\vskip.1cm \item  The payoff depends on the actions of the co-players as well as on the frequencies of interactions. Both quantities depend on the probability state of the system.

\vskip.1cm \item Interactions are nonlocal and nonlinearly additive in a way that the dynamics of a few entities do not lead straightforwardly to the dynamics of the whole system.

\end{itemize}

A qualitative description of phenomenological examples of interactions is as follows:

\begin{enumerate}

 \item  \textit{Competitive (dissent):} One of the interacting particles increases its status by taking advantage of the other, obliging the latter to decrease it. Competition brings advantage to only one of the two.

\vskip.1cm  \item   \textit{Cooperative (consensus):} The interacting particles exchange their status,  one by increasing it and the other one by decreasing it. Interacting active particles show a trend to share their micro-state.

  \vskip.1cm  \item    \textit{Learning:} One of the two modifies, independently from the other, the micro-state. It learns by reducing the distance between them.

\vskip.1cm  \item \textit{Hiding-chasing:} One of the two attempts to increase the overall distance from the other, which attempts to reduce it.

\vskip.1cm  \item   \textit{Mixed competitive-cooperative:} A-particles do not share the same dynamics, but some of them act competitively and some cooperatively.
\end{enumerate}

If the dynamics of interaction depends on space the following geometrical quantities, and related properties, must be introduced:

\begin{itemize}

  \item \textit{Visibility domain:}  $\Omega_v$ which is the domain within which an a-particle can visualize the presence of the other a-particles.

 \vskip.2truecm   \item  \textit{Sensitivity domain:} $\Omega_s$ which is the domain within which an a-particle can visualize the presence of the other a-particles. If $\Omega_s \subseteq \Omega_v$, then interactions occur within  $\Omega_s$.  If $\Omega_v \subseteq \Omega_s$, then interactions occur within  $\Omega_v$. The interaction domain $\Omega$ is defined as the intersection between  $\Omega_v$ and $\Omega_s$.

\vskip.2truecm   \item  \textit{The size of the sensitivity domain} $\Omega_s$ depends on the amount of information which can be obtained by an active particle, hence  $\Omega_s$ depend on the distribution function. The theory proposed in~\cite{[BCC08]} suggests that the size of $\Omega_s$ depends on a critical density, namely a critical number of particles. The mathematical formalization in~\cite{[BH17]} indicates how  $\Omega_s$ is related to the velocity direction and visibility angle of each a-particle.

\end{itemize}

The following active particles are supposed to be involved, for each functional subsystem, in the interactions:

\vskip.1cm \no  $\bullet$ \textit{Test} particles of the $i$-th functional subsystem with microscopic state, at time $t$,
delivered by the variable $(\bx, \bv, \bu)$, whose distribution function is $f_i = f_i(t, \bx, \bv, \bu)$.
The test particle is assumed to be representative of the whole system.

\vskip.1cm  \no  $\bullet$ \textit{Field} particles of the $k$-th functional subsystem with microscopic state, at time $t$,
defined by the variable $(\bx^*, \bv^*, \bu^*)$, whose distribution function is $f_k = f_k(t, \bx^*, \bv^*, \bu^*)$.

\vskip.1cm  \no  $\bullet$ \textit{Candidate} particles, of the $h$-th functional subsystem,  with microscopic state, at time $t$,
defined by the variable $(\bx_*, \bv_*, \bu_*)$, whose distribution function is $f_h = f_h(t, \bx_*, \bv_*, \bu_*)$.

\vskip.1cm   Let us now consider short range interactions, when particles interact within an interaction domain $\Omega$
 generally small with respect to the domain $\Sigma$ containing the whole system; and let us use the term $i$-particle  to denote a particle in the $i$--th functional subsystem. Bearing in mind that a precise definition and computing of $\Omega$ still needs to be given, the theory states that the modeling of interactions is delivered by the following quantities:

\vskip.1cm \noindent  $\bullet$ \textit{Interaction rate for conservative dynamics}: $\eta_{hk}[\f](\bx, \bv_*, \bu_*, \bx^*, \bv^*,\bu^*)$ which models the frequency of the interactions between a candidate $h$-particle with state $(\bx, \bv_*, \bu_*)$ and a field $k$-particle with
state $ (\bx^*, \bv^*, \bu^*)$. Analogous expression is used for interactions between test and field particles.

\vskip.1cm \noindent  $\bullet$ \textit{Interaction rate for non-conservative dynamics}: $\mu_{hk}[\f](\bx, \bv_*, \bu_*, \bx^*, \bv^*,\bu^*)$ which is analogous to $\eta_{hk}$, but corresponding to  proliferative and destructive interactions.

\vskip.1cm   \noindent  $\bullet$ \textit{Transition probability density}: $\cC_{hk}^i[\f](\bv_*\rightarrow \bv,\, \bu_* \rightarrow \bu \,|\, \bx, \bv_*,\bu_*,\bx^*, \bv^*, \bu^*)$ which denotes  the probability density that a  candidate $h$-particle, with state $(\bx, \bv_*, \bu_*)$, ends up into the state of the test particle of the  $i$-th FS after an interaction  with a  field $k$-particle.

\vskip.1cm  \noindent  $\bullet$  \textit{Proliferative term}: $\cP_{hk}^{i} [\f](\bv_* \rightarrow \bv, \, \bu_*\rightarrow  \bu\,|\, \bx, \bv_*, \bu_*,\bx^*,\bv^*, \bu^*)$ which models the proliferative events for a candidate $h$-particle, with state $(\bx, \bv_*, \bu_*)$, into the $i$-th functional subsystem after interaction  with a field $k$-particle with state $(\bx^*, \bv^*, \bu^*)$.

\vskip.1cm  \noindent  $\bullet$ \textit{Destructive term}: $\cD_{ik} [\f](\bx, \bv, \bu, \bx^*, \bv^*, \bu^*)$ which models  the rate of destruction for a test $i$-particle  in its own functional subsystem after an interaction  with a field $k$-particle with state $(\bx^*, \bv^*, \bu^*)$.

\vskip.1cm
These quantities can be viewed in terms of rates by multiplying their interaction rate with the terms modeling
transition, proliferative, and destructive events. Hence we have  \textit{transition rate:}  $\eta_{hk}[\f] \, \cC_{hk}^i[\f]$,  \textit{proliferation rate:}  $\mu_{hk}[\f] \, \cP_{hk}^{i} [\f]$, and  \textit{destruction rate:} $\mu_{ik}[\f] \, \cD_{ik} [\f]$.

Mathematical structures are obtained by number balance of a-particles within an elementary volume of the space of microscopic states, mechanics and activity, of particles
\begin{eqnarray*}
&&\hbox{\bf Variation rate of the number of active particles}\\
&& \hskip.5truecm = \hbox{\sl Inlet flux rate  by conservative interactions}\\
&& \hskip1truecm - \hbox{\sl Outlet flux rate  by conservative interactions}\\
&& \hskip1.5truecm +  \hbox{\sl Inlet flux rate  by proliferative  interactions and mutations}\\
&& \hskip2truecm - \hbox{\sl Outlet flux rate by destructive interactions and mutations},
\end{eqnarray*}
where the inlet flux includes the dynamics of mutations.

This balance relation  corresponds to the following general structure:
\begin{equation}\label{general}
\left(\p_t  +  \bv \cdot \p_\bx \right)\, f_i(t, \bx, \bv, \bu) = \big(\cC_i -  \cL_i  + \cP_i -  \cD_i\big)[\f](t, \bx, \bv, \bu),
  \end{equation}
where the various terms $\cC_i, \cL_i, \cP_i$ and $\cD_i$  can be formally expressed, consistently with the definition of the interaction terms.

\vskip.1cm \noindent \textbf{Remark 3.1.} \textit{A commonly applied assumption is that the terms $\cP_{hk}^{i}$ and  $\cD_{ik}$ depend, in addition to $\f$, only on the activity variables, namely $\cP_{hk}^{i}[\f](\bu_* \to \bu|\bu_*, \bu^*)$ and $\cD_{ik}[\f](\bu, \bu^*)$. This assumption is used in the equations below.}
\vskip.1cm
In the spatially homogeneous case, the  mathematical structure is specialized as follows:
\begin{eqnarray}\label{homogen}
&& \p_t f_i(t, \bu) =\bigg(\cC_i [\f] - \cL_i[\f]  + \cP_i [\f]  - \cD_i[\f]\bigg) (t, \bu) \nonumber \\
&& \hskip.1cm =   \sum_{h,k=1}^n \,\int_{D_\bu \times D_\bu} \eta_{hk}[\f] (\bu_*,\bu^*)\, \mathcal{C}_{hk}^i[\f]\left(\bu_* \to \bu|\bu_*,\bu^* \right)f_h(t,\bu_*)f_k(t,\bu^*)\,d\bu_*\,  d\bu^*\nonumber \\
&& \hskip1cm  -  f_i(t,\bu) \sum_{k=1}^n \, \int_{D_\bu}  \eta_{ik}[\f] (\bu,\bu^*)\, f_k(t,\bu^*)\,d\bu^* \nonumber \\
&& \hskip.1cm  +  \sum_{h,k=1}^n \,\int_{D_\bu \times D_\bu} \mu_{hk}[\f](\bu_*,\bu^*) \mathcal{P}_{hk}^i[\f]\left(\bu_* \to \bu|\bu_*,\bu^* \right) \, f_h(t,\bu_*)f_k(t,\bu^*)\, d\bu_*\, d\bu^* \nonumber \\
&& \hskip1cm  -  f_i(t,\bu) \sum_{k=1}^n \,\int_{D_\bu} \mu_{ik}[\f](\bu,\bu^*)\, \mathcal{D}_{ik}(\bu,\bu^*) f_k(t,\bu^*)\,d\bu^*.
\end{eqnarray}

The same calculations, in the  spatially inhomogeneous case, correspond to Eq.~(\ref{general}), where the interaction terms are given by:
\begin{eqnarray}
\cC_i[f] &=&\sum_{h,k=1}^n  \int_{\Omega \times D_\bu \times D_\bu \times D_\bv \times D_\bv} \eta_{hk}[\f](\bw_*,\bw^*) \, \cC_{hk}^i[\f] (\bv_* \to \bv,\,  \bu_* \to  \bu |\bw_*, \bw^*) \nonumber \\
  &{}& \hskip1cm  \times f_h(t, \bx, \bv_*, \bu_*) f_k(t, \bx^*, \bv^*,
\bu^*)\, d\bx^*\,d\bv_*\ d\bv^*\,d\bu_*\, d\bu^*,
 \end{eqnarray}
 \begin{equation}
\cL_i[f] =f_i(t, \bx, \bv, \bu) \sum_{k=1}^n  \int_{\Omega\times D_\bv\times D_\bu} \eta_{ik}[\f](\bw,\bw^*)\,
f_k(t, \bx^*, \bv^*, \bu^*)\, d\bx^* d\bv^* \, d\bu^*,
\end{equation}
\begin{eqnarray}
\cP_i[f] &=&\sum_{h,k=1}^n  \int_{\Omega \times D_\bu \times D_\bu \times D_\bv \times D_\bv} \mu_{hk}[\f](\bw_*,\bw^*)\, \mathcal{P}_{hk}^i[\f]\left(\bu_* \to \bu|\bu_*,\bu^* \right) \nonumber\\
 &{}& \hskip1cm \times f_h(t, \bx, \bv_*, \bu_*)f_k(t, \bx^*, \bv^*, \bu^*)\, d\bx^*\, d\bv_*\,d\bv^*\,d\bu_*\, d\bu^*,
\end{eqnarray}
\begin{eqnarray}\label{death}
\cD_i[f] &=& f_i(t, \bx, \bv, \bu) \sum_{k=1}^n  \int_{\Omega \times D_\bv\times D_\bu} \mu_{ik}[\f](\bw,\bw^*)\,\cD_{ik}[\f](\bu,\bu^*)\nonumber\\
 &{}& \hskip1cm \times f_k(t, \bx^*, \bv^*, \bu^*)\, d\bx^*\, d\bv^* \, d\bu^*,
\end{eqnarray}
 where $\bw$, $\bw_*$ and $\bw^*$  denote the microscopic states $(\bx,\bv,\bu)$, $(\bx,\bv_*,\bu_*)$ and $(\bx^*,\bv^*,\bu^*)$, respectively. Detailed calculations, which are not repeated here, indicate how $\Omega$ can be computed when the sensibility area $\Omega_s$ is given by an arc of circle, with radius $R_s$ around the velocity direction. Then, if the visibility arc, symmetric or non-symmetric, is known,  $R_s$ is referred to the critical number of a-particles necessary to a sufficient information.

Further developments of the structures (\ref{general})--(\ref{death}) will be outlined in the next subsection referring to a critical analysis on the limits and possible developments of these structures. Here, we just anticipate some technical remarks that can contribute to enlighten the properties of these mathematical structures in view of derivation of models:
\begin{enumerate}

\item  The use of distribution functions, rather than probability density, as dependent variables, accounts for a dynamics with  a variable number of a-particles due to birth and loss processes.

\vskip.1cm \item  Mutations can be modeled by birth processes which can generate entities (gain) more fitted to the environment, who in turn might generate new entities again more fitted to the outer environment. Selection can be modeled by the death (loss) of entities less fitted to the environment.

\vskip.1truecm   \item Two types of interactions are taken into account: \textit{micro-micro} or \textit{micro-macro}, where the term  \textit{macro} corresponds to macroscopic quantities obtained by weighted averaging of the distribution function.

\end{enumerate}

Mathematical models can be obtained by a detailed specialization of these mathematical structures, namely by modeling the interaction terms $\Omega, \eta, \mu, \cC, \cP, \cD$. However, some common features of the modeling approach can be given by a qualitative description to be formalized for each case study, as we shall see in the next sections.

\vskip.1cm $\bullet$  \textit{Hierarchy:} The study of human crowds~\cite{[BGO19]} has suggested a hierarchy in the decision making by which walkers develop their walking strategies, namely by interactions which firstly induce modification of the social state, subsequently walkers modify their walking direction, and finally they adapt the speed to the local new flow direction.

\vskip.1cm  $\bullet$  \textit{Sensitivity domain:} This quantity is defined by a cone with vertex in $\bx_i$, vertex angle $\Theta$, and with axis along with the velocity. $\Omega_i$ is finite being truncated at a distance $R$ which might be related to the critical finite number of active particles which have a sensitive influence.  The sensitivity domain might be modified by visibility problems.

\vskip.1cm  $\bullet$   \textit{Interaction rate:} The modeling of the interaction rates can be referred to a distance between the interacting entities by a metric suitable to account both for the distance between the interacting entities and that of their statistical distribution.

\vskip.1cm  $\bullet$   \textit{Social action:} The social action depends on the interaction of each particle with those in its sensitivity domain. It is specific of each system to be modeled.

\vskip.1cm  $\bullet$  \textit{Mechanical action:} Mechanical actions follow the rules of classical mechanics, but the parameters leading to accelerations depend on the social state by models to be properly defined.

\subsection{Critical analysis and conceptual alternatives}\label{subsec:3.2}

It can be rapidly observed that the structures  (\ref{general})--(\ref{death}) can lead to the derivation of mathematical models once the various interaction terms $\eta, \mu, \cC, \cP, \cD$ can be modeled on the basis of a phenomenological-theoretical interpretation of each specific system object of the modeling approach. However, a detailed analysis of the structures (\ref{general})--(\ref{death}) is necessary to verify the ability
to capture the complexity features identified within the answer to the key question KQ2 within the modeling framework visualized in the flow chart in Figure 1.

Bearing all above in the mind let us focus on each of the key features and analyze critically how far the mathematical theory can effectively account for them. The study of this problem refers also to specific applications.

\vskip.1cm \noindent  $\bullet$  \textit{Ability to express a strategy:} This ability  is modeled by the activity variable, say \textit{behavioral ``soft'' variable}, then all components can have a reciprocal influence. If it includes both behavioral and mechanical variables, then the latter is influenced by the former. An example of the first case is given by the dynamics of idiosyncratic learning which affects the skill in market sharing~\cite{[BDKV20]}. An example of the latter case appears in crowd dynamics~\cite{[BGO19]} as the strategy by which walkers move depends on the emotional state of people in the crowd.

\vskip.1cm \noindent $\bullet$  \textit{Heterogeneity:}  The use of the distribution function over the activity as the independent variable accounts for the heterogeneous behavior of a-particle. Hence players are probability distributions and the interactions account for this specific feature.

\vskip.1cm \noindent   $\bullet$   \textit{Nonlinearity of  interactions:} The modeling of interactions generally leads to nonlinearly additive outputs from interactions. The output can also depend on the distribution functions. For instance, models of opinion formation include the sensitivity of a-particles not only to individual a-particles, but also to first order moments. This type of dynamics characterizes, as an example, aggregation into
political opinions as individual attitudes are modified not only by individual based interactions, but also by individuals as a whole.

  \vskip.1cm \noindent    $\bullet$  \textit{Learning ability:} Individual entities, namely a-particles, learn from past experience~\cite{[BDG16],[BDG16C],[BGO17]}. As a consequence, the dynamics of interaction is modified by the level of learning which is heterogeneously acquired by each individual. Besides the application treated in Section 4, an example is given by the reciprocal learning and hiding between security forces and criminality, where learning acts simultaneously with hiding followed, respectively by namely criminal and chasing actions~\cite{[BCKS15]}.

 \vskip.1cm \noindent   $\bullet$  \textit{Darwinian mutations and selection:} All living systems are evolutionary, as birth processes can generate entities more fitted to the environment. These, in turn, generate new  entities again more fitted to the outer environment. An immediate application appears in the immune competition in cancer dynamics, where several mutations generate cancer cells~\cite{[HW11],[HART01],[WEI07]}, while the immune system evolves by learning to produce selection~\cite{[MF19]}.

\vskip.1cm

Although we have verified that the mathematical structures produced by the kinetic theory for active particles can capture the complexity features of living systems, additional key problems need to be considered to validate the mathematical theory. We have selected the following ones: \textit{large deviations} that might lead to unpredictable events, specifically the so-called \textit{black swan}, \textit{aggregation into groups of affinity} which generate endogenous networks, \textit{continuity assumption of the distribution function}, and \textit{interactions of the different dynamics depicted by the activity variable} should be accounted for based on Section 2.  Further critical analysis is proposed in the next section, while some of the aforementioned problems, and various others, can be treated referring to applications such as those reported in Section 5.  An additional problem, which is a stone guest in all applications, is the search for a \textit{multiscale vision} which is the key passage towards a mathematical theory of living systems.

\section{On a mathematical theory of behavioral swarms}\label{sec:4}

The kinetic theory for active particles requires the continuity assumption of the distribution function deemed to describe the collective state of the system. An alternative approach has been introduced in~\cite{[BHO20]} by inserting the activity variable in the micro-state of the interacting entities whose collective  dynamics is modeled within a pseudo-Newtonian framework. This generalization has generated a new class of swarms models which can be viewed as a development of the celebrated Cucker and Smale model~\cite{[CS07]} which generated a vast literature on the modeling, qualitative analysis, and computational applications. The interested reader is referred to the survey~\cite{[ALBI19]}, where  the theory of swarms is  specifically treated in Sections 5 and 6.

The objective of this section consists in showing how the pioneering ideas of~\cite{[BHO20]} can be developed to extend the achievements of this paper to the modeling of living systems along the strategy proposed in Section 2. We firstly introduce the theory proposed in\cite{[BHO20]}, which is somehow inspired by some applications of the theory to  social and economical problems~\cite{[ABH13],[BCLY17],[BCK19],[BDKMT20]} and, subsequently, we tackle  the aforementioned developments into more general frameworks.

\subsection{Mathematical structures of behavioral swarm theory}\label{subsec:4.1}

Let us consider  the dynamics of a heterogeneous  swarm  of $N$  interacting self-propelled particles, which can be subdivided into $m$ functional subsystems, each $k$ functional subsystem contains $n_k$ active particles. Heterogeneity refers to the ability of each particle to express a  social state which will be called \textit{activity}. Individual particles are labeled by the subscript $ik$ and the individual state of each $ik$-th particle is defined by position $\bx_{ik}$, velocity $\bv_{ik}$ and activity $u_{ik}$, while the speed of the activity variable, namely the time derivative of $u_{ik}$ is denoted by $z_{ik}$.

The velocity  $\bv_{ik}$  can be represented in polar coordinates as follows:
\begin{equation}
\bv_{ik} = \{v_{ik}, \bomega_{ik}\}, \hskip.5cm  v_{ik} \in[0,1],  \hskip.5cm \bomega_{ik} = \frac{\bv_{ik}}{||\bv_{ik}||}\va
\end{equation}
where the unit vector  $\bomega_{ik}$ can be represented by the angular variables referred to a  cartesian system and where the cartesian components of the  position $\bx_{ik}$ are referred to the characteristic length $\ell$ of the system, while $v_{ik}$ has been referred to the limit velocity $v_M$ which can be reached by the fastest particle.

In more details, if the system is localized in a bounded domain $\Sigma$, the positive constant $\ell$ is the diameter of the circle containing $\Sigma$, while if the system moves  in an unbounded domain, $\ell$ is simply referred to the domain  $\Sigma_0$  containing the particles at $t=0$. The sets of all positions, velocities, and activities of the whole system are denoted by $\bx$, $\bv$ and $\bu$, respectively. The activity variable $u_{ik}$ takes values in $[0,1]$, where the limit values $u_{ik}=0$ and $u_{ik}=1$ represent, respectively, the lowest and highest values of the activity. We firstly consider the simpler case of a scalar activity variable, which simplified notations, while the more general case of vector variables are studied in the next sections.

As in the case of the kinetic theory approach, we look for the derivation of a general mathematical structure consistent with the complexity features of living systems. This structure is deemed to provide the conceptual framework for the derivation of models. This  conceptual approach should refer to complexity features analogous to those considered in the KTAP approach:
\begin{enumerate}
\item Each particle is able to develop a specific strategy which is heterogeneously distributed.

\vskip.1cm \item Interactions can be nonlocal and  nonlinearly additive.

\vskip.1cm \item A decisional hierarchy can be used by assuming that interactions first modify the activity and subsequently the motion which depends also on the activity.

\vskip.1cm \item  Each a-particle has a sensitivity domain and interacts with all particles within the said domain.

\vskip.1cm \item The approach is such that firstly a mathematical structure is derived  and subsequently it is implemented by models of interactions which lead to specific models.

\end{enumerate}

We consider interactions which modify the microscopic state and might also generate proliferative destructive events. We  consider the specific quantities deemed  to model, still at a formal level, interactions:

\vskip.1cm \noindent  $\bullet$  $\eta_{ik}$ models the interaction rate of individual based interactions between ${ik}$-particle with all particles in the sensitivity domain.

\vskip.1cm \noindent   $\bullet$  $\Omega_{ik}$ is the vision-based sensitivity domain of the $ik$-particle.

\vskip.1cm \noindent   $\bullet$ $\vf_{ik}$ denotes the action, which occurs with the rate  $\eta_{ik}$,  over the activity variable over the ${ik}$-particle by all particles in $\Omega_{ik}$.

\vskip.1cm \noindent  $\bullet$  $\psi_{ik}$ denotes the acceleration, which occurs with the rate  $\eta_{ik}$,  over the mechanical  variable by all particles in $\Omega_{ik}$.

\vskip.1cm  \noindent  $\bullet$  $\cP_{ik}$  and $\cD_{ik}$ model, respectively, the destruction and  the proliferation of $ik$-particles, which occurs with the rate  $\eta_{ik}$,  due to the action  by all particles in $\Omega_{ik}$.

\vskip.1cm  \noindent  $\bullet$ $n_k= n_k(t)$ denotes number, which depends on time, of particles in each k-FS.

\vskip.1cm  Formal calculations and obvious interpretation of notations yield, to the following \textit{second order framework for a mixture of functional subsystems}:
\begin{equation}\label{multi-swarm-s}
\begin{cases}
\displaystyle  \frac{d\bx_{ik}}{dt} =  \bv_{ik}, \\[2mm]
\displaystyle \frac{d\bv_{ik}}{d t} = \sum_{k=1}^m \sum_{{jq} \in \Omega_{ik}} \eta_{ik}(\bx_{ik}, \bv_{ik}, \bu_{ik}, \bz_{ik}, \bx_{jq}, \bv_{jq}, \bu_{jq}, \bz_{jq})  \\[1mm]
\hskip3cm \times\, \psi_{ik}(\bx_{ik}, \bv_{ik}, \bu_{ik}, \bz_{ik}, \bx_{jq}, \bv_{jq}, \bu_{jq}, \bz_{jq}), \\[2mm]
\displaystyle \frac{d u_{ik}}{dt} =  z_{ik}, \\[2mm]
\displaystyle \frac{d z_{ik}}{d t} = \sum_{k=1}^m  \sum_{{jq} \in \Omega_{ik}} \eta_{ik}(\bx_{ik}, \bv_{ik}, \bu_{ik}, \bz_{ik}, \bx_{jq}, \bv_{jq}, \bu_{jq}, \bz_{jq})  \\[1mm]
\hskip3cm \times\, \vf_{ik}(\bx_{ik}, \bv_{ik}, \bu_{ik}, \bz_{ik}, \bx_{jq}, \bv_{jq}, \bu_{jq}, \bz_{jq}),\\[2mm]
\displaystyle \frac{d n_{k}}{d t} =  \sum_{k=1}^m  \sum_{{jq} \in \Omega_{ik}} \eta_{ik}(\bx_{ik}, \bv_{ik}, \bu_{ik}, \bz_{ik}, \bx_{jq}, \bv_{jq}, \bu_{jq}, \bz_{jq})  \\[1mm]
\hskip3cm \times\, \left(\cP_{ik} - \cD_{ik}\right)(\bx_{ik}, \bv_{ik}, \bu_{ik}, \bz_{ik}, \bx_{jq}, \bv_{jq}, \bu_{jq}, \bz_{jq}),
  \end{cases}
\end{equation}
where all components of the dependent variables, which play a role in the dynamics, appear in the argument of the interaction terms.

\subsection{Critical analysis towards further developments}\label{subsec:4.2}

The concise presentation delivered in the preceding sections is consistent with the lectures~\cite{[BDKV20]}, specifically with Lecture 6 which shows how the theory can be applied to modeling the dynamics of prices in an open market, where buyers and producers interact. In addition, the same lecture shows how the theory can model a variety of behavioral swarms where individuals firstly exchange their social state and subsequently they develop their dynamics conditioned from the time-evolving social state.

However, we cannot yet state that a complete theory is available as important topics should still be developed. Specifically, we refer to the dynamics across functional subsystems as well as mutations and selections. This specific dynamics is important non only in the modeling of the dynamics of biological systems, for instance multicellular systems in cancer phenomena~\cite{[WEI07]} and the immune competition\cite{[MF19]}, where active particles mutate and can be selected, but also in social-economical systems, where pseudo-Darwinist dynamics are observed in various cases.

A further topic, which deserves attention, is the modeling in network dynamics focused on exogenous and endogenous networks. In the former case, in addition to migration across nodes,  interactions should include not only interactions within each node, but also between active particles in the nodes and the nodes of the network viewed as a whole. The modeling approach, proposed in~\cite{[KNO13],[KNO14]} within the framework of the KTAP theory,
should be technically extended to the case of the theory of swarms.

Analogous reasonings can be focused on endogenous networks, where nodes are created by spontaneous aggregation of active particles, generally based on affinity dynamics. Therefore, an additional dynamics should be inserted in the overall model. Some guide principles towards modeling of endogenous networks have been given in~\cite{[DLO17]} and~\cite{[ABCK]}, referring to the  KTAP theory, but additional work should still be developed towards a complete theory.

\section{Applications looking ahead to modeling perspectives}\label{sec:5}

This section presents a concise review and critical analysis of applications to modeling of different types of behavioral systems based on the mathematical theory proposed in the first part of this paper.  As we have seen, mathematical models are derived by inserting a mathematical description of interactions at the micro-scale into a differential structure selected among those presented in Sections 3 and 4.
 We present, in the following subsections, some selected applications based on the kinetic theory methods and that of behavioral swarms. The selection is specifically related to Lectures~\cite{[BBDG20]} and it goes beyond~\cite{[BBGO17],[BKS13]} as  the modeling of interactions can depend not only on the micro-state of the interacting entities, but also on the probability state of the interacting active particles. Moreover, interactions can be selective and non-symmetric. The term \textit{nonlinear interactions} is used to denote both these specific typology of interactions.

Our review is limited to very recent years and specifically to  paper published in the last five years and it is limited to  models where the dynamics of  interactions is nonlinear as specified by the aforementioned definition. A huge enlargement of literature would appear by including interactions whose output would depend only on the micro-state of the interacting pairs.

\subsection{Modeling collective learning dynamics}\label{subsec:5.1}
The definition \textit{collective learning} is used to denote the dynamics by which an individual learns a well defined skill from a population of interacting individuals.  In more details,  the following definition can guide the modeling approach~\cite{[CAP99]}:
\begin{quote}
\textit{It is a social process of cumulative knowledge, based on a set of shared rules and procedures which allow individuals to coordinate their
actions in search for problem solutions.}
\end{quote}
 Collective learning is an \textit{interactive process}, where the  transfer of knowledge is induced by interactions which occurs in the individual's mind as a social and participatory process and that increases his mental knowledge~\cite{[PIAGET76],[VYG78]}. In addition, it is \textit{lastly cumulative}, since it accumulates over time. A specific example is the \textit{social learning}, originated with the development of psychology sciences,  which occurs when the individual learns new behaviors (as specific skills) and concepts from others~\cite{[BAN89],[SP98]}.

Methods of the kinetic theory of active particles on this specific dynamics have been proposed in~\cite{[BDG16]} and~\cite{[BGO17]} with application to modeling scholar learning. Further developments have been treated in~\cite{[BD19]},  where the modeling of the interaction between learning and different dynamics has been investigated. This paper refers to the most important research perspectives which aim at investigating that different types of learning are combined with other dynamics and have an immediate influence on the rules by which the said dynamics develops.
This topic is treated in Lecture 3 of~\cite{[BBDG20]}.

\subsection{Vehicular traffic and human crowds}\label{subsec:5.2}

Modeling and mathematical problems of vehicular traffic and human crowds were recently reviewed in~\cite{[ALBI19]}. Therefore, we will avoid overlap by mentioning that  the pioneering path on vehicular traffic  were traced in~\cite{[PH71],[PF75]}. The  modeling of interactions by density dependent rules have been introduced in~\cite{[CDF07]}  for discrete velocity models under the assumption that the speed of vehicles could attain only a finite number of velocities defined of a grid depending on the local density and in~\cite{[DT07]} in the case of a fixed grid. Further developments of this pioneering idea have been given in~\cite{[FT13],[FT15]}, while a qualitative and computational analysis of the initial value problem has been developed in~\cite{[BDF12]}.

The modeling of human crowds accounting for nonlinear interactions has been initiated in~\cite{[BBK13]} and further developed by various authors as reported in~\cite{[ALBI19]}, see also~\cite{[EAA19],[EA19]}. Recent studies were firstly focused on the modeling of the dynamics of social interactions~\cite{[BGO19],[BRSW15]} and subsequently to virus  contagion problems~\cite{[KQ19],[KQ20]}. These two topics can be viewed as key research perspective to be developed within the multiscale framework proposed in~\cite{[ABGR20]} for the modeling of human crowds. Lecture 4 of~\cite{[BBDG20]} focuses on some of the topics described in the above concise report.

\subsection{From the immune competition to modeling virus pandemics}
\label{subsec:5.3}
 Modeling the immune competition between cancer and immune cells ended up with the first class of models derived within a mathematical theory of active particles. The main results on this topic can be found in the book~\cite{[BD06]}. This research line has been further developed in~\cite{[BDK13]} to account for mutations and selection of tumor cells as well as for the learning ability of the immune system. The modeling of the role of macrophages has been developed in~\cite{[EG20]}. The book~\cite{[MF19]} is a useful reference to understand the complex dynamics in immunology, while the book~\cite{[WEI07]} provides a precious description of cancer biology.

 The pioneering paper~\cite{[DDS09]}, devoted to the modeling of virus mutations followed by a learning dynamics,  provides  some ideas which be developed towards a modeling approach to depict the complex mutation-learning dynamics specifically referred to  COVID-19. Actually, the modeling of the dynamics of the corona-virus requires, as shown in~\cite{[BBC20]}, a multiscale approach suitable to  go far beyond deterministic population dynamics as individual reactions to the infection and pandemic events are heterogeneously distributed throughout the population. In fact, the contagion occurs at the high scale of individuals depending on the viral charge inside each  individual whose  dynamics is at smaller scales determined  by the competition between virus particles and the immune system. In addition,  spatial dynamics and interactions are important features to be included in the modeling approach as the dynamics are generated by nonlocal interactions and transportation devices as shown in Lecture 4 of~\cite{[BBDG20]}.

\subsection{From behavioral to evolutionary economy}\label{subsec:5.4}

The very first application to modeling social dynamics by the kinetic theory methods arguably belongs to the pioneering paper~\cite{[JS92]}, while the kinetic theory of active particles was applied  in~\cite{[BD04]} by models of interactions related to the micro-state only with focus on opinion formation. Modeling the interaction of more than one dynamics accounting for nonlinear interactions has been studied in~\cite{[BHT13]}, where it has been shown that unfear social dynamics enhances radicalization and opposition to governments, and in~\cite{[BCKS15]} modeling the complex interactions between security forces and criminality by a paper which enlighten the role of training of security forces as the key feature to contrast criminality. Further studies have followed in the field of behavioral and political economy, as examples~\cite{[DKLM17],[DL14],[DL15],[DLO17]}. This research activity refers to the stream of behavioral economy~\cite{[Thaler16],[TS09]}.

Recent research activity has been devoted to evolutionary economy  based on the interaction between mathematics and economy specifically focused on  particles and firms meeting within the framework of capitalist economies. In more details,  we refer to the dynamics of
the capitalist system which is characterized by processes of endogenous self-sustained growth, punctuated by small and big crises. The following statement, from page 83 of~\cite{[Schumpeter]},  enlightens this concept:
\begin{quote}
 \textit{Industrial mutation -- if I may use the biological term -- that incessantly revolutionizes the economic structure from within, incessantly destroying the old one, incessantly creating a new one. This process of Creative Destruction is the essential fact about capitalism.}
\end{quote}

In modern capitalism, business firms are a central locus of the efforts to advance technologies, develop new products and operate new production processes~\cite{[DOSI84]}. The key features, see~\cite{[BDKV20],[DV17],[DPV17]}  of the dynamics are as follows:

\vskip.1cm \noindent -- \textit{Persistently heterogeneous firms are nested in competitive environments, which shape their individual economic fate and, collectively, the evolution of the forms of industrial organization.}

 \vskip.1cm \noindent  -- \textit{The process through which heterogeneous firms compete, let us call it Schumpeterian competition, on the basis of the products and services they offer and obviously their prices, and get selected -  with  some firms growing, some declining, some going out of business, some new ones always entering.}

  \vskip.1cm \noindent  -- \textit{Such processes of competition and selection are continuously fueled by the activities of innovation, adaptation, imitation by incumbent firms and by entrants.}
 \vskip.1cm
 In more details, the approach developed in~\cite{[BDKV20]} includes the following dynamics:  \textit{Learning} or empirically the \textit{within effect}: idiosyncratic innovation, imitation, changes in technique of production; and \textit{selection} or empirically the \textit{between effect}: market interaction where more competitive firms gain at the expense of less competitive ones. The kinetic theory approach accounts for two functional subsystems which are nested into a hierarchical structure:
\begin{enumerate}
\item \textbf{Subsystem 1: Evolutionary landscape:} It represents the dynamics of learning to which firms are subject to. It is meant to capture the arrival of new technologies, new ideas, new organizational practices. It evolves \textit{independently} from firm interactions, and it follows a \textit{continuous growth} process. In economic terms, it represents the evolution of the technological frontier.

    \vskip.1cm \item \textbf{Subsystem 2: Endogenous system of interactions:} It comprises two distinct levels of interactions: one which determines the advancement of knowledge of each individual firm through the action of the first Subsystem, the second which entails the competition in the market arena among heterogeneous firms in terms of knowledge level.
\end{enumerate}

The key objective consists in developing the first mathematical step developed in~\cite{[BDKV20]} into a consistent mathematical theory of evolutionary systems in economy.

\section{Reasonings on multiscale frameworks}\label{sec:6}

Modeling always needs a \textit{multiscale approach}, where the dynamics at the large scale needs to be properly related to the dynamics at the low scales. For instance, in biology the functions expressed by  a cell are determined by the dynamics at the molecular (genetic) level. Some of the models reviewed in this section include not only \textit{micro-scale interactions},  but also \textit{micro-macro interactions} occur between particles and FSs viewed as a whole being represented by their mean value. As an example we can refer again to the~\cite{[BBDG20]} which presents a multiscale vision of a virus pandemic in a complex world. It is shown how the contagion propagates in a crowd where the awareness of the contagion is heterogeneously distributed and subsequently is followed by a within host dynamics. The scenarios of hospitalization, recovery and death can be delivered by mathematical models  developed within a multiscale framework, where the dynamics of individuals depends on the dynamics at smaller scales inside each individual by the competition between virus particles and the immune system.

General,  \textit{multiscale methods} are developed by  modeling of individual based interactions, which are used  firstly to derive models at the microscopic scale, while subsequently, these micro-scale models are used to derive kinetic type models, namely at the mesoscopic scale. The third step consists in developing asymptotic methods which lead to hydrodynamical models by writing equations in a dimensionless form to work out a small dimensionless parameters related to the mean distance between particles. Macroscopic equations are derived by letting this parameter to zero under suitable assumptions. This micro-macro derivation corresponds to the sixth Hilbert problem posed by David Hilbert for classical particles in physics. A development to active particles has been proposed in~\cite{[BC17],[BC19]}.

Finally, we return to the concept of \textit{interdisciplinary approach} which has pervaded the whole paper and, specifically, our lectures~\cite{[BBDG20]}. We do believe that the  \textit{interdisciplinary vision} of science is not simply an approach, but a new way of thinking research activity. This concept is precisely the message in~\cite{[Greco]}

\input{References-KTAP-20-12}

\end{document}

%% file: References-KTAP-20-12.tex
%
%
%